\documentclass[prc,superscriptaddress,unsortedaddress,twocolumn,showpacs]{revtex4}
\usepackage{amsmath}
\usepackage{amssymb}
\usepackage{times}
\usepackage{mathrsfs}
\usepackage{multirow}
\usepackage{bm}
\usepackage{wrapft}
\usepackage{color}
\usepackage{graphicx}
\usepackage{ulem}

\def\fun#1#2{\lower3.6pt\vbox{\baselineskip0pt\lineskip.9pt
\ialign{$\mathsurround=0pt#1\hfil##\hfil$\crcr#2\crcr\sim\crcr}}}

\newcommand{\beq}{\begin{equation}}
\newcommand{\eeq}{\end{equation}}
\newcommand{\bea}{\begin{eqnarray}}
\newcommand{\eea}{\end{eqnarray}}

\newcommand{\bfi}[1]{\mbox{\boldmath $#1$}}
\newcommand{\bfis}[1]{\mbox{\boldmath ${\scriptstyle #1}$}}

\renewcommand{\vr}{{\bfi r}}
\newcommand{\vR}{{\bfi R}}

\newcommand{\viR}{{\bfis R}}

\begin{document}

\title{Reexamination of microscopic optical potentials based on multiple scattering theory}

\author{Kosho Minomo}
\email[]{minomo@rcnp.osaka-u.ac.jp}
\affiliation{Research Center for Nuclear Physics, Osaka University, Ibaraki 567-0047, Japan}

\author{Kouhei Washiyama}
\affiliation{Center for Computational Sciences, University of Tsukuba, Tsukuba 305-8577, Japan}

\author{Kazuyuki Ogata}
\affiliation{Research Center for Nuclear Physics, Osaka University, Ibaraki 567-0047, Japan}

\date{\today}

\begin{abstract}
\noindent
{\bf Background:} 
Microscopic optical potentials have been successful in describing
nucleon-nucleus and nucleus-nucleus scattering. Some essential
ingredients of the framework, however, have not been examined
in detail.\\
{\bf Purpose:}
Applicability of the microscopic folding model is systematically investigated.
Effect of an antisymmetrization factor (ASF) appearing in multiple scattering
theory, theoretical uncertainty regarding the local density approximation (LDA),
and the validity of a prescription for nonlocality, the Brieva-Rook (BR) localization,
of the microscopic potential, are quantitatively estimated for nucleon-nucleus scattering;
investigation on the ASF is carried out for also deuteron-nucleus
scattering.\\
{\bf Methods:}
A single folding model with the Melbourne $g$-matrix interaction
and the SLy4 Skyrme-type Hartree-Fock-Bogoliubiv (SLy4-HFB) density
is employed
for evaluating a nucleon-nucleus microscopic optical potential.
Deuteron-nucleus scattering is described by the
continuum-discretized coupled-channels method incorporating
the microscopic proton-nucleus and neutron-nucleus potentials.\\
{\bf Results:} 
The ASF is found to affect proton total reaction cross sections
$\sigma_{\rm R}$
for a $^{12}$C target below 200 MeV by about 10\%.
Effect of the ASF on $\sigma_{\rm R}$ is negligibly small
if a target nucleus is heavy or scattering energy is above 200~MeV;
elastic cross sections are hardly affected by the ASF for all the
reaction systems considered.
Below 65 MeV,
still the BR localization works quite well. However, at energies
below about 50~MeV, the LDA becomes less accurate for evaluating
elastic cross sections at backward angles. This is the case also
for $\sigma_{\rm R}$ of $p$-$^{12}$C below about 200~MeV.\\
{\bf Conclusions:} 
The microscopic model is applicable to nucleon-nucleus scattering
above 25 MeV for target nuclei in a wide range of mass numbers.
Deviation of
calculated results from experimental data is less than about 10\%.
\end{abstract}


\maketitle

\section{Introduction}
\label{sec1}
Construction of optical potentials, which describe a one-body potential between a projectile and a target nucleus for elastic scattering,
is one of the most fundamental subjects in nuclear reaction theory.
For nucleon-nucleus scattering, some global optical potentials have been constructed phenomenologically
by means of the rich data of differential cross sections, total reaction cross sections, and spin observables.
Koning and Delaroche proposed a global optical potential for proton and neutron scattering up to 200 MeV in a traditional way,
in which the local Woods-Saxon and its derivative forms are assumed~\cite{kon03}.
The Dirac phenomenology provides precise optical potentials for proton scattering~\cite{ham90,coo93}, at higher incident energies in particular.
Very recently, nonlocal optical potentials, which well reproduce the measured data even at very low energies,
have been constructed with the dispersive optical model~\cite{dic16}.
Thus, reliable optical potentials were obtained for proton-nucleus elastic scattering as far as the data are available.
Optical potentials are applied to also analysis of other reaction processes, for example, inelastic scattering, breakup, and transfer reactions.

Recently, microscopic approach to optical potentials took large steps.
The $g$-matrix folding model is one of the most practical methods to construct an optical potential.
In the model, the folding potential is obtained by folding a $g$-matrix interaction,
which is an effective nucleon-nucleon interaction in nuclear matter, with a target density.
Basically, $g$-matrix interactions are obtained by solving the Bethe-Bruckner-Goldstone equation.
Up to now, many types of $g$-matrix interactions have been developed, for example,
JLM~\cite{jeu77}, M3Y and its modified versions (DDM3Y and CDM3Y)~\cite{ber77,far85,kho97}, CEG~\cite{yam83}, and Melbourne~\cite{amo00} are often used.

As a noteworthy achievement, the folding calculation with the Melbourne $g$-matrix interaction can reproduce
the measured cross sections for nucleon-nucleus scattering with no free adjustable parameter~\cite{amo00,deb05}.
In addition to the conventional nucleon-nucleon interactions,
modern interactions including three-body-force (3BF) effects have been developed~\cite{fur08,yam13,raf13,toy15}.
Those interactions are also used to investigate the 3BF effects for nucleus-nucleus scattering~\cite{min16,fur16} and proton knockout reactions~\cite{min17a};
the 3BF effects are found not to be essential for nucleon-nucleus elastic scattering.
Furthermore, great efforts have been done to derive optical potentials by {\it ab initio} approach,
though it is still restricted to reaction systems containing a very limited number of nucleons~\cite{rot17}.
Inspired by the pioneering and state-of-the-art microscopic studies by
the Melbourne group, several works have been reported for microscopically
describing not only nucleon-nucleus scattering but also nucleus-nucleus
scattering and other reaction processes. In some cases three-body or four-body reaction models
incorporating microscopic distorting potentials are employed~\cite{kik13,min14,neo16,min17b}.
In these studies, single- or double-folding model with the Melbourne
$g$-matrix interaction was adopted, with the local density approximation
(LDA) and the localization prescription proposed by Brieva and Rook~\cite{BR77,BR77a,BR78},
that is, the BR localization.
However, applicability of this microscopic framework has not been
investigated systematically, at energies below 65~MeV and above 200~MeV
in particular. Discussion on the role of an antisymmetrization factor (ASF)
(see Sec.~\ref{sec2}) appearing in multiple scattering theory is also
missing. Furthermore, the applicability of the LDA and the BR localization
has been examined only in very limited cases.
It should be noted that in the original work by the Melbourne group
nonlocal terms have explicitly been treated without using the BR localization.

In this study we carry out systematic studies on the applicability of
the folding model with the LDA and the BR localization. Properties of
its ingredients, that is, the ASF, theoretical uncertainty regarding
the LDA, and the validity of the BR localization, are investigated.
We mainly focus on proton-nucleus scattering; the role of the ASF is
discussed also for deuteron-nucleus scattering by means of the
continuum-discretized coupled-channels method (CDCC)~\cite{kam86,aus87,yah12}.

The contents of this paper are organized as follows.
In Sec.~\ref{sec2}, we show a theoretical framework and the Schrodinger equation to be solved.
The results of nucleon elastic scattering and total reaction cross sections in Secs.~\ref{sec3a} and \ref{sec3b}, respectively.
In Sec.~\ref{sec3c}, we evaluate the theoretical ambiguity of LDA in the $g$-matrix folding model,
and in Sec.~\ref{sec3d}, we check the nonlocality of the microscopic optical potential.
Application of the microscopic optical potentials for deuteron scattering is shown in Sec.~\ref{sec3e}.
Finally, Sec.~\ref{sec4} is devoted to a summary.

\section{Theoretical framework}
\label{sec2}

\subsection{Schr{\"o}dinger equation with an effective interaction}
\label{sec2a}
We consider nucleon (N) scattering from a target nucleus (T) with mass number $A_{\rm T}$.
A theoretical foundation of the $g$-matrix folding model is given by multiple scattering theory~\cite{ker59}.
We start with the following Schr{\"o}dinger equation described by a bare nucleon-nucleon interaction between N and a $j$-th nucleon in T, $v_{j}$,
\bea
\bigg[K+h_{\rm T}+\sum_{j\in{\rm T}}v_{j}-E\bigg]\Psi=0.
\label{bare}
\eea
Here, $E$ is the total energy, $K$ is the kinetic energy operator between N and T,
$h_{\rm T}$ is the internal Hamiltonian of T.
$\Psi$ is the total wave function.
Multiple scattering theory allows one to solve the following equation with an effective interaction $\tau_{j}$,
\bea
\bigg[K+h_{\rm T}+\frac{A_{\rm T}-1}{A_{\rm T}}\sum_{j\in{\rm T}}\tau_{j}-E\bigg]\hat{\Psi}=0,
\label{tau}
\eea
instead of solving Eq.~(\ref{bare}).
The transition matrix $T$ for elastic scattering is given by
\bea
T=\frac{A_{\rm T}}{A_{\rm T}-1}T',
\label{tmat}
\eea
where $T'$ is the transition matrix obtained by solving Eq.~(\ref{tau}).
$(A_{\rm T}-1)/A_{\rm T}$ in Eq.~(\ref{tau}) and $A_{\rm T}/(A_{\rm T}-1)$ in Eq.~(\ref{tmat}) appear as a result of antisymmetrization in multiple scattering theory;
below we call each of them an antisymmetrization factor (ASF).

With the nuclear matter approximation for implementing the medium effects in $\tau_{j}$, it is replaced with a $g$-matrix interaction.
Multiplying Eq.~(\ref{tau}) by a target ground-state wave function from the left,
a one-body Schr{\"o}dinger equation for the relative motion between N and T for elastic scattering is derived:
\bea
\bigg[K_{\viR}+\frac{A_{\rm T}-1}{A_{\rm T}}U(\vR)+U_{\rm Coul}(\vR)-E_{\rm c.m.}\bigg]\chi(\vR)=0.
\label{Schr}
\eea
Here, $U(\vR)$ is a localized folding potential between N and T, which consists of the central and spin-orbit parts; the explicit form is given in Ref.~\cite{toy13}.
$U_{\rm Coul}$ is the Coulomb potential and $E_{\rm c.m.}$ is the incident energy in the center-of-mass frame.

The total reaction cross section $\sigma_{\rm R}^{}$ is given by
\bea
\sigma_{\rm R}^{}=\frac{\pi}{k^2}\sum_{L,J}(2J+1)(1-|S_{LJ}|^2),
\eea
where
\bea
S_{LJ}=-\frac{1}{A_{\rm T}-1}+\frac{A_{\rm T}}{A_{\rm T}-1}S'_{LJ}.
\eea
Here, $S'_{LJ}$ is the scattering matrix obtained by solving Eq.~(\ref{Schr}),
and $k$ is the relative wave number between N and T.
$L$ ($J$) is the orbital (total) angular momentum appeared in the partial-wave decomposition of $\chi$.

The ASFs have been ignored in most cases. However, these may play an important role when lighter targets are involved.
It should be noted that no ASF appears when a phenomenological optical potential is considered.

\subsection{Three-body model}
\label{sec2b}
In Refs.~\cite{neo16,min17b} multiple scattering theory was applied to
deuteron scattering. The weakly bound property of deuteron allows one
to adopt a three-body reaction model and the Schr\"odinger equation
is given by
\bea
\left[K+h_{d}+\frac{A-1}{A}(U_{p{\rm T}}+U_{n{\rm T}})-E_{\rm c.m.}\right]\chi=0.
\label{Schr_deu}
\eea
Here, $h_{d}$ is the internal Hamiltonian of deuteron and
$U_{p{\rm T}}$ ($U_{n{\rm T}}$) is a distorting potential between proton (neutron) and the target obtained microscopically.
$A$ is the product of the mass numbers of deuteron and target nucleus,
that is, $A=2A_{\rm T}$.
In the previous studies~\cite{neo16,min17b} the ASF $(A-1)/A$ was
disregarded, validity of which is examined in Sec.~\ref{sec3e}.
The three-body equation (\ref{Schr_deu}) is solved by using the continuum-discretized coupled-channels method (CDCC)~\cite{kam86,aus87,yah12}.

\subsection{Skyrme Hartree-Fock-Bogoliubov density}
\label{sec2c}
The one-body densities used in our microscopic optical potential
are calculated within the Hartree-Fock-Bogoliubov (HFB) model
with SLy4 Skyrme energy density functionals~\cite{cha98}.
In the pairing channel, we use mixed-type pairing functionals
with a quasiparticle cutoff of 60 MeV.
The pairing strength is determined so as to reproduce
the neutron pairing gap of 1.25 MeV in ${}^{120}$Sn.
We use the computer code \textsc{lenteur}~\cite{ben},
where one-body HFB equations are solved in spherical symmetry
and time-reversal symmetry.
Odd and odd-odd nuclei are calculated by constraining
average particle numbers
to its odd numbers without breaking time-reversal symmetry.

\section{Results and Discussion}
\label{sec3}
In this paper, we calculate proton scattering on $^{12}$C, $^{40}$Ca, and $^{208}$Pb from 25 to 800 MeV.
For comparison, we adopt two types of target densities.
One is SLy4-HFB density as explained in Sec.~\ref{sec3c},
and the other is the phenomenological one deduced from the analysis of electron scattering~\cite{vri87}.
For the latter, the proton density is obtained by unfolding the charge distribution with the finite-size effect of proton charge~\cite{sin78}, and
the neutron densities for $^{12}$C and $^{40}$Ca are approximated to be the same as those of proton;
for $^{208}$Pb, we adopt the neutron density distribution deduced from the analysis of proton elastic scattering~\cite{zen10}.
We refer the phenomenological density as the Sum-of-Gaussian (SOG) density below.

We use the Melbourne $g$-matrix interaction~\cite{amo00}.
At energies higher than 300~MeV, we adopt the $g$-matrices with the optical model prescription (OMP)~\cite{ger98} that
modify the bare interaction to implement the meson production effects.
The relativistic correction for the kinematics is considered in solving Eq.~(\ref{Schr}), that is,
the wave number is estimated according to the relativistic kinematics and we take the reduced energy instead of the reduced mass.

\subsection{Nucleon elastic scattering}
\label{sec3a}
\begin{figure}[tbp]
\begin{center}
\includegraphics[width=0.4\textwidth,clip]{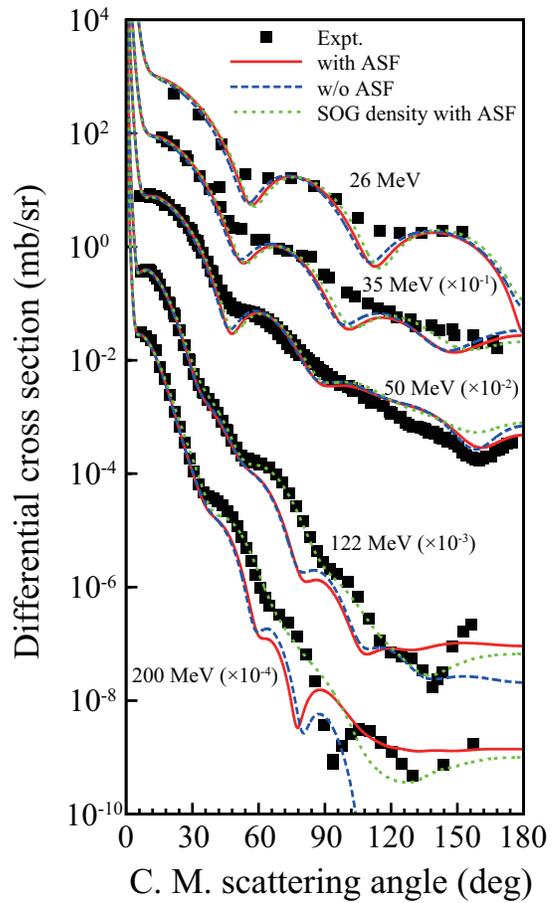}
\caption{(Color online)
Differential cross sections for $p$-$^{12}$C elastic scattering as a function of the scattering angle in the center-of-mass frame.
The solid and dashed lines show the results obtained with the microscopic density with and without the ASFs, respectively, whereas
the dotted lines correspond to the result with the SOG density and the ASF.
The incident energies are 26, 35, 50, 122, and 200 MeV from the top to the bottom, and
the results are scaled for visibility by the factors in parentheses near the curves.
The experimental data are taken from Refs.~\cite{har99,fab80,fan67,cla71,rus71,com80,mey83,mey81}.
}
\label{fig1}
\end{center}
\end{figure}

Figure~\ref{fig1} shows the differential cross sections for $p$-$^{12}$C elastic scattering from 26 to 200 MeV.
The solid (dashed) lines correspond to the results using the microscopic density with (without) the ASF.
The dotted lines represents the result with the SOG density and the ASF.
It is found that the ASF shifts the diffraction pattern to backward angles very slightly.
This effect becomes smaller as the incident energy increases since
$(A_{\rm T}-1)/A_{\rm T}$ in Eq.~(\ref{tau}) and $A_{\rm T}/(A_{\rm T}-1)$ in Eq.~(\ref{tmat}) are almost canceled.
This indicates multistep processes in terms of $U$ become less important at higher energies, as expected in Ref.~\cite{ker59}.

Comparing the results with the two types of densities, the difference is negligibly small at low energies, whereas it appear at large angles
at higher energies.
Agreement between the theoretical results and the experimental data are satisfactorily well at 26, 35, and 50~MeV, except around the dips. At higher energies, the dotted lines retain this feature, whereas the other two show some deviation at backward angles This indicates that $^{12}$C is too light for the SLy4-HFB calculation to describe its density distribution, in the nuclear interior region in particular. It will be worth pointing out that even in this situation the cross sections calculated with the SLy4-HFB density agree well with the data at forward angles, in which the cross sections are dominantly large.

\begin{figure}[tbp]
\begin{center}
\includegraphics[width=0.4\textwidth,clip]{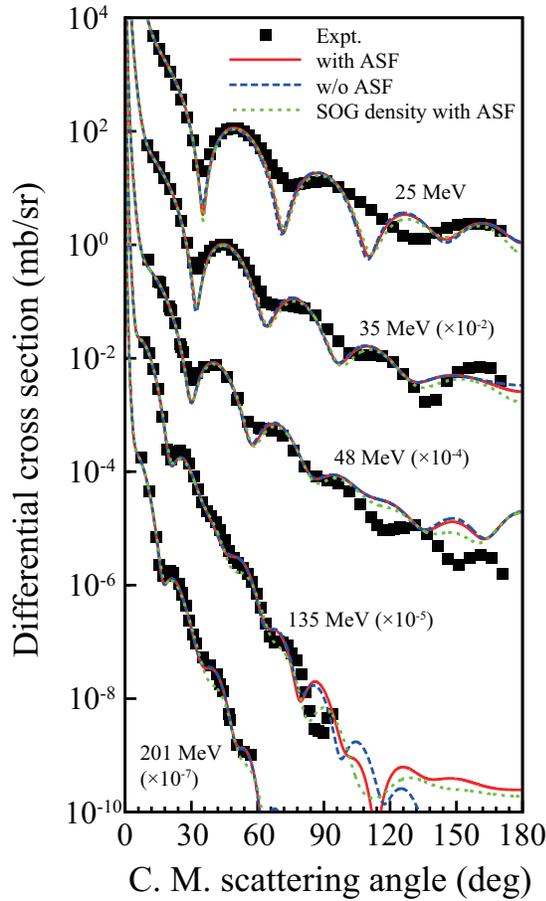}
\caption{(Color online)
Differential cross sections for $p$-$^{40}$Ca elastic scattering.
The meaning of the lines is same as that in Fig.~\ref{fig1}.
The experimental data are taken from Refs.~\cite{mcc86,nad81,sei93}.
}
\label{fig2}
\end{center}
\end{figure}

\begin{figure}[tbp]
\begin{center}
\includegraphics[width=0.4\textwidth,clip]{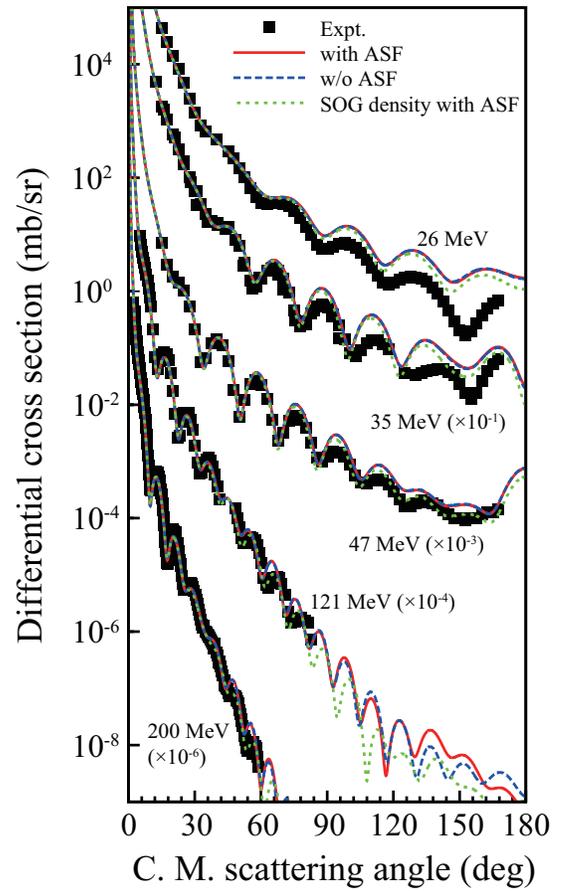}
\caption{(Color online)
Differential cross sections for $p$-$^{208}$Pb elastic scattering.
The meaning of the lines is same as that in Fig.~\ref{fig1}.
The experimental data are taken from Refs.~\cite{oer74,wag75,nad81,hut88}.
}
\label{fig3}
\end{center}
\end{figure}

Features of the results for $^{40}$Ca and $^{208}$Pb targets, shown in
Figs.~\ref{fig2} and \ref{fig3} respectively, are quite similar to those
in Fig.~\ref{fig1} except that i) the ASF little affect the results and ii)
the two densities give almost the same cross sections even at high energies.
Consequently, the results with the SLy4-HFB density reproduce the experimental
data very well; at larger angles for scattering on $^{208}$Pb still the
results with the SOG density gives a slightly better agreement.
The overshooting at backward angles at low energies in Fig.~\ref{fig3}
may be due to the Coulomb excitation that is not taken into account in
the present calculation.

\begin{figure}[tbp]
\begin{center}
\includegraphics[width=0.4\textwidth,clip]{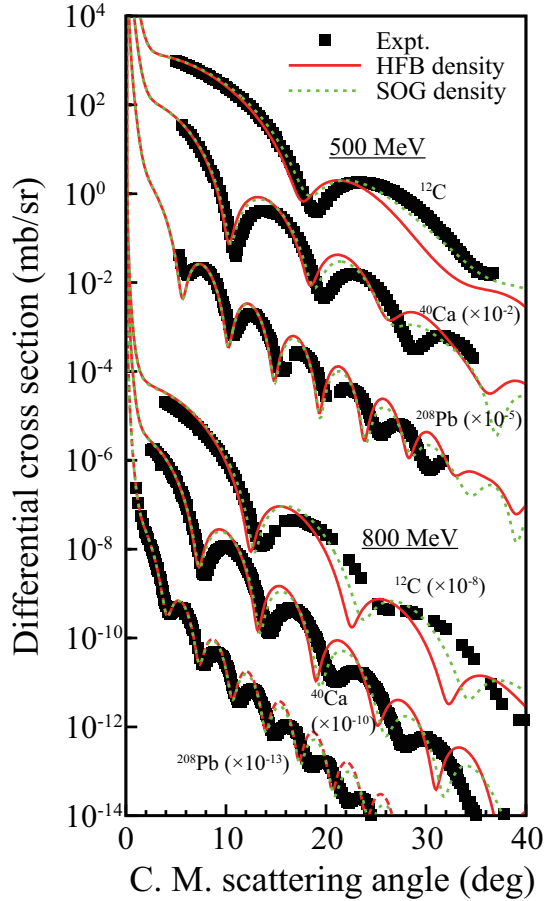}
\caption{(Color online)
Differential cross sections for proton elastic scattering from $^{12}$C, $^{40}$Ca, and $^{208}$Pb at high energies.
The incident energies of the three results from the top correspond to around 500 MeV,
and the lower three to 800 MeV.
The solid (dotted) line shows the result with the SLy4-HFB (SOG) density including the ASF.
The experimental data are taken from Refs.~\cite{hof90,hof88,hof81,bla81,ray81,ble82,hof80}.
}
\label{fig4}
\end{center}
\end{figure}

Figure~\ref{fig4} shows the differential cross sections for proton elastic scattering
from $^{12}$C, $^{40}$Ca, and $^{208}$Pb at 500 and 800 MeV.
Results with the SLy4-HFB density neglecting the ASF are found to
completely agree with those with the ASF and thus not plotted.
Although the agreement between the theoretical and experimental results is not bad at very forward angles,
the diffraction pattern of the theoretical calculations shifts to forward angles compared with that of the data.
The SOG density gives slightly better agreement with the experimental data but still it somewhat deviates from the data.
This may suggest a limitation of the nonrelativistic approach to high energy scattering.

\subsection{Total reaction cross sections for nucleon scattering}
\label{sec3b}
\begin{figure}[tbp]
\begin{center}
\includegraphics[width=0.45\textwidth,clip]{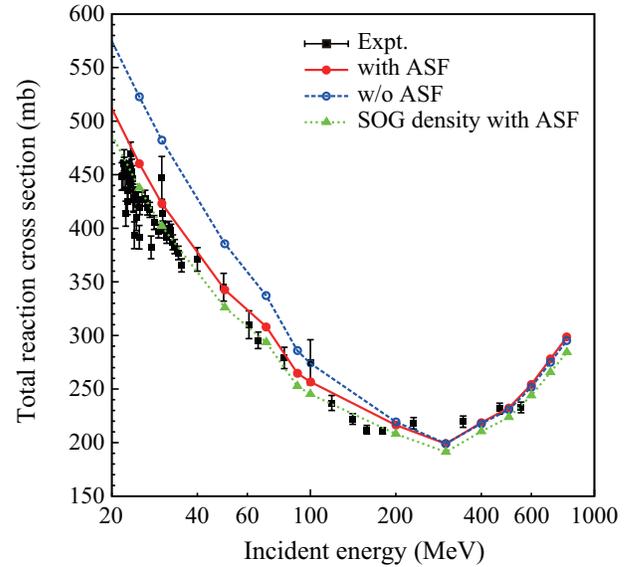}
\caption{(Color online)
Energy dependence of the total reaction cross sections for $p$-$^{12}$C scattering.
The closed (open) circles correspond to the result with (without) the ASF and
the triangles to the results with the SOG density including the ASF.
The lines are guides for eyes.
The experimental data are taken from Refs.~\cite{sla75,men71,ing99,auc05,tay61,ren72}.
}
\label{fig5}
\end{center}
\end{figure}

Figure~\ref{fig5} shows the total reaction cross sections $\sigma_{\rm R}$ for $p$-$^{12}$C scattering,
as a function of the incident energy.
The closed (open) circles correspond to the results with (without) the ASF and
the triangles to the results with the SOG density including the ASF. The lines are guides for eyes.
The experimental data are taken from Refs.~\cite{sla75,men71,ing99,auc05,tay61,ren72}.

At low energies, there is a 10\% difference between the results with and without the ASF,
and the former shows a better agreement with the experimental data.
This difference disappears at energies higher than 200~MeV.
It is found that $\sigma_{\rm R}$ calculated with the SOG density is smaller than that with the SLy4-HFB density by several percent.
This difference reflects the larger matter RMS radius of the Sly4-HFB density than that of the SOG density.

\begin{figure}[tbp]
\begin{center}
\includegraphics[width=0.45\textwidth,clip]{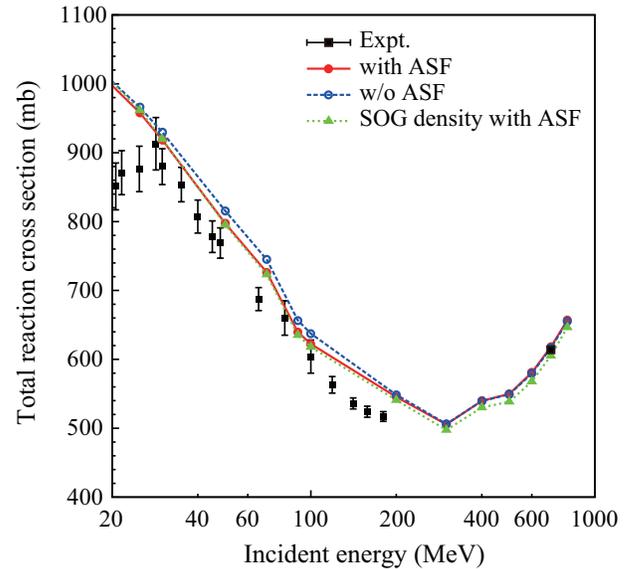}
\caption{(Color online)
Same as Fig.~\ref{fig5} but for $p$-$^{40}$Ca.
The experimental data are taken from Refs.~\cite{dic70,car75,tun64,ing99,auc05,and79}.
}
\label{fig6}
\end{center}
\end{figure}

\begin{figure}[tbp]
\begin{center}
\includegraphics[width=0.45\textwidth,clip]{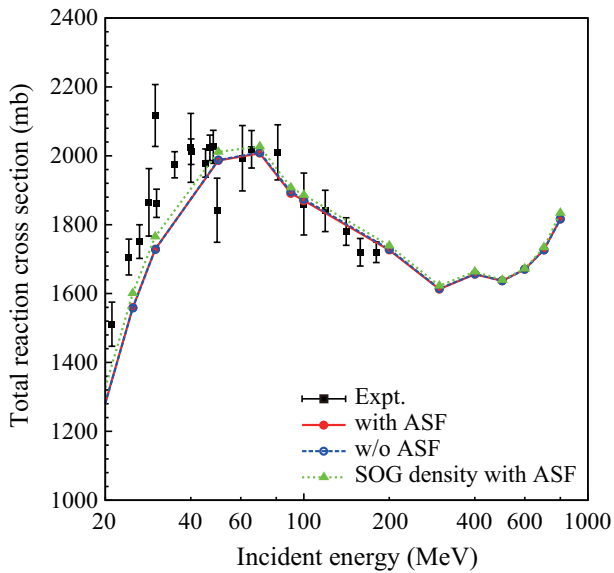}
\caption{(Color online)
Same as Fig.~\ref{fig5} but for $p$-$^{208}$Pb.
The experimental data are taken from Refs.~\cite{car75,tun64,men71,ing99,auc05}.
}
\label{fig7}
\end{center}
\end{figure}

Figures~\ref{fig6} and \ref{fig7} show $\sigma_{\rm R}$ for
$p$-$^{40}$Ca and $p$-$^{208}$Pb scattering, respectively.
For the former, the effect of the ASF is significantly smaller than in
Fig.~\ref{fig5}, whereas for the latter the effect is totally negligible.
In both cases, the two densities give almost the same results.
$\sigma_{\rm R}$ for $p$-$^{40}$Ca slightly overshoots the experimental data.
For $p$-$^{208}$Pb the agreement with the data is quite satisfactory except
the undershooting below about 50~MeV.


\subsection{Theoretical ambiguity coming from the local density approximation}
\label{sec3c}
The $g$-matrix interaction is applied to finite nuclei with the local density approximation (LDA),
that is, we choose the density at a certain point, $\vr_g$, as the density-dependence of the $g$-matrix interaction in the folding procedure.
We have three choices for $\vr_g$:
i) the mid point of the interacting two nucleons ($\vr_m$),
ii) the coordinate of the internal nucleon ($\vr$), and iii) that of the incoming nucleon ($\vR$).
If the LDA is completely valid, the calculated results are independent of the choice of $\vr_g$.
By comparing the results with ii) and iii), which correspond to the two extreme cases, we can estimate how large the theoretical ambiguity of the LDA is at most.
In this paper we consider the scattering below 65~MeV; for energies above 65~MeV, see Ref.~\cite{min10}.

\begin{figure}[tbp]
\begin{center}
\includegraphics[width=0.45\textwidth,clip]{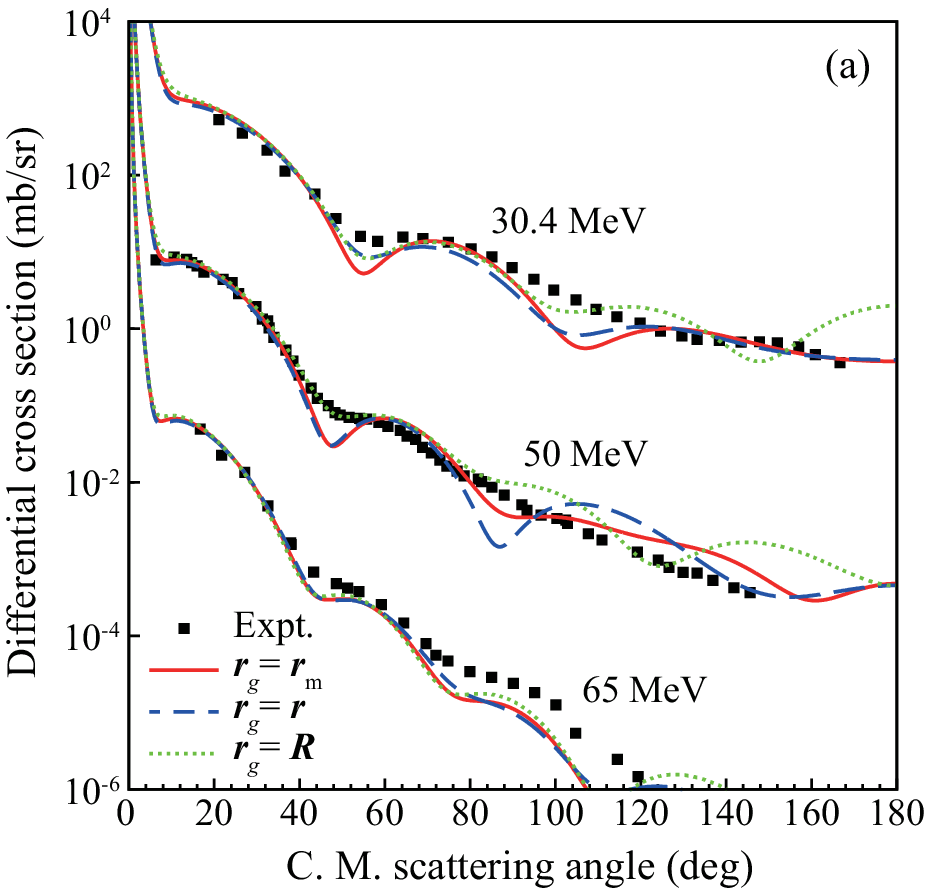}
\vspace{10mm}
\includegraphics[width=0.45\textwidth,clip]{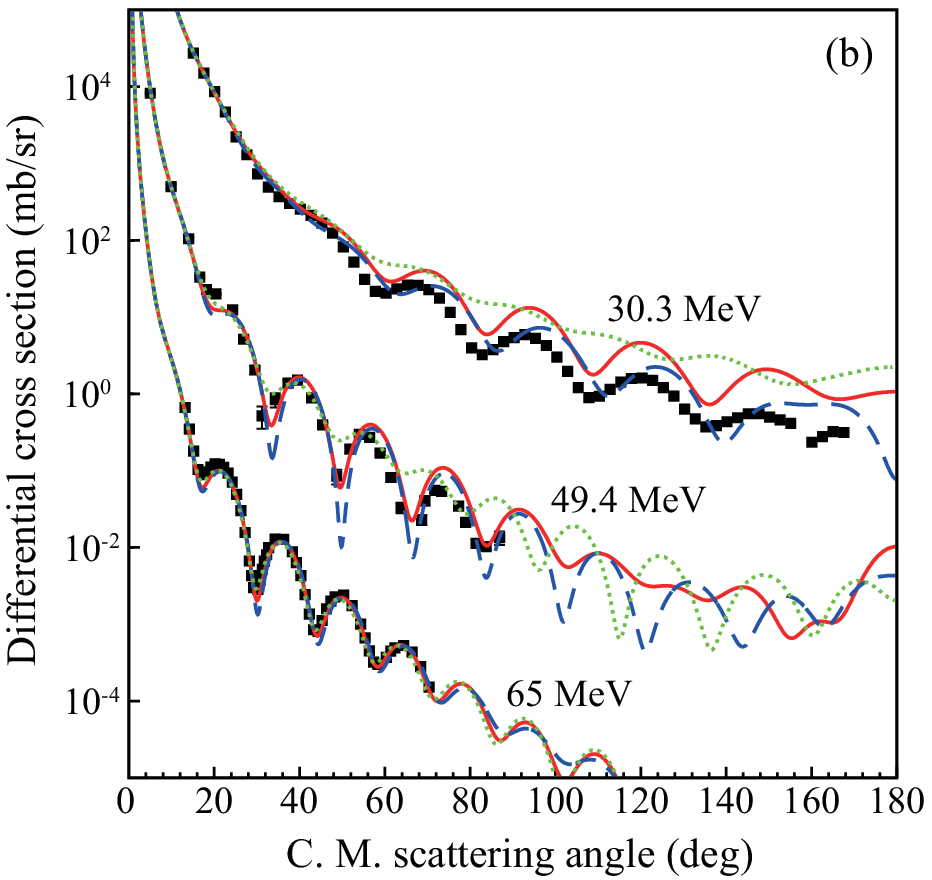}
\caption{(Color online)
Differential cross sections for a) $p$-$^{12}$C and b) $p$-$^{208}$Pb scattering
calculated with the different choice of $\vr_g$.
The horizontal axis is the scattering angle in the center-of-mass frame.
The solid, dashed, and dotted lines correspond to i) $\vr_g=\vr_m$, ii) $\vr_g=\vr$, and iii) $\vr_g=\vR$, respectively.
From the top, the incident energies correspond to 30.4, 50, 65 MeV in panel (a),
and 30.3, 49.4, 65 MeV in panel (b).
The results at higher energies are scaled for visibility.
The experimental data are taken from Refs.~\cite{gre72,rus71,kat85,oer74,man71,sak82}.
}
\label{fig8}
\end{center}
\end{figure}

Figure~\ref{fig8} shows the differential cross sections for $p$-$^{12}$C and $p$-$^{208}$Pb scattering
calculated with the different choices of $\vr_g$.
The solid, dashed, and dotted lines correspond to i) $\vr_g=\vr_m$, ii) $\vr_g=\vr$, and iii) $\vr_g=\vR$, respectively.
The experimental data are taken from Refs.~\cite{gre72,rus71,kat85,oer74,man71,sak82}.
below 50 MeV, the diffraction pattern varies by changing $\vr_g$.
Features of the results seem not to depend on the target nuclei.
At 65 MeV, the density-dependence of the $g$-matrix itself becomes relatively weak so that
the three lines agree quite well each other; this is consistent with the finding in Ref.~\cite{min10} above 65~MeV.

\begin{figure}[tbp]
\begin{center}
\includegraphics[width=0.45\textwidth,clip]{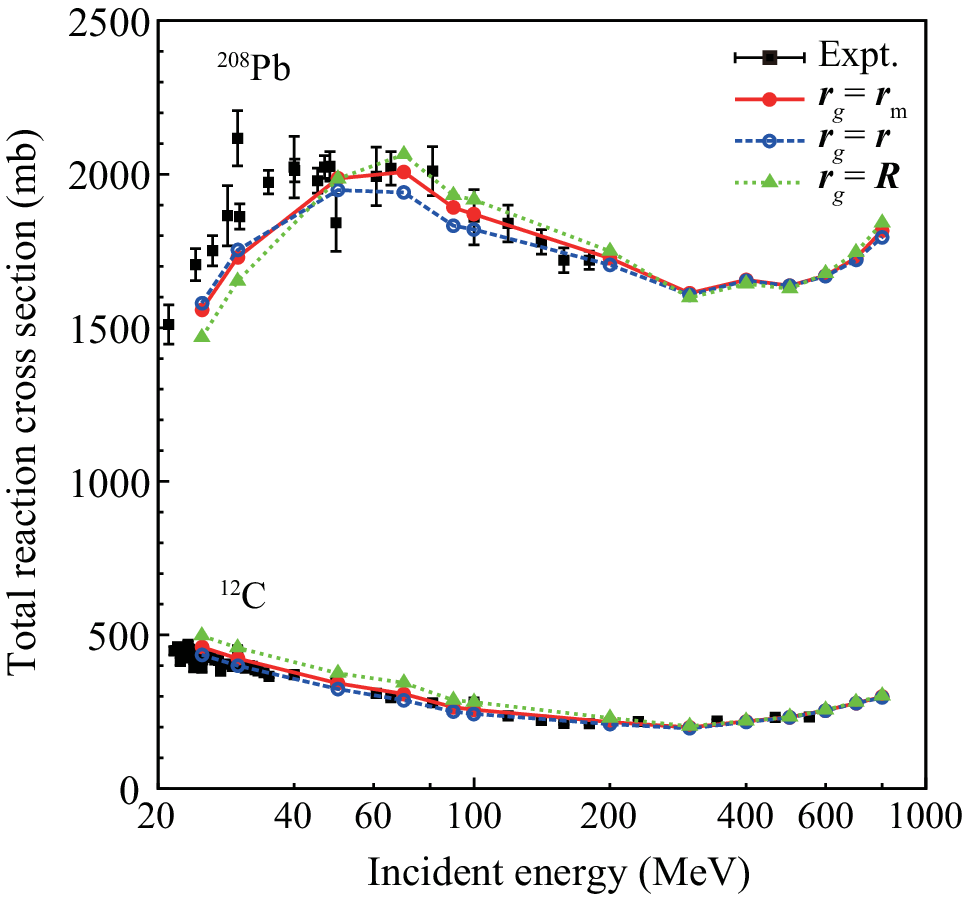}
\caption{(Color online)
The energy dependence of the total reaction cross sections of
$p$-$^{12}$C and $p$-$^{208}$Pb scattering.
The open circles, the closed circles, and the triangles correspond to
choices i), ii), and iii) for $\vr_g$, respectively.
The experimental data are same as in Figs.~\ref{fig5} and \ref{fig7}.
}
\label{fig9}
\end{center}
\end{figure}

A similar investigation on $\sigma_{\rm R}$ for
$p$-$^{12}$C and $p$-$^{208}$Pb is shown in Fig.~\ref{fig9}.
The open circles, the closed circles, and the triangles represent the
results corresponding to choices i), ii), and iii), respectively.
For each system, the three results are different from each other
by several percent below 200~MeV. The small but finite difference
will indicate the theoretical uncertainty from the use of the LDA,
even though the uncertainty may be overestimated in the present
analysis.

\subsection{Treatment of nonlocality}
\label{sec3d}
\begin{figure}[tbp]
\begin{center}
\includegraphics[width=0.45\textwidth,clip]{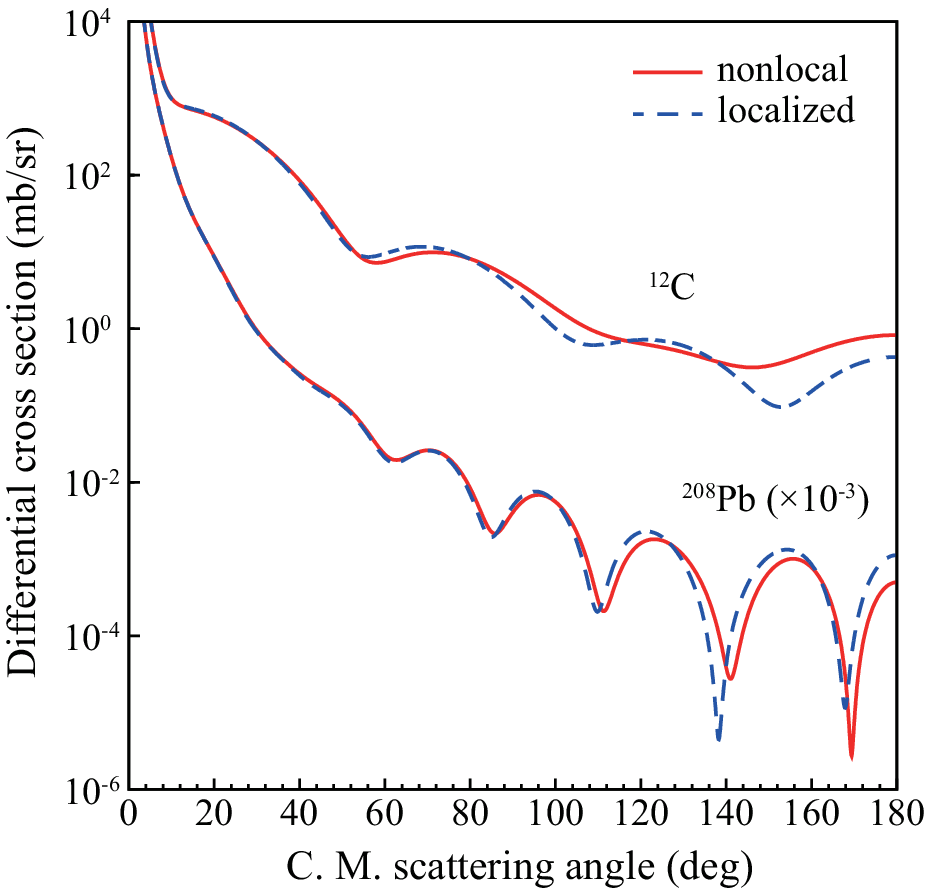}
\caption{(Color online)
Differential cross sections for $p$-$^{12}$C and $p$-$^{208}$Pb at 30 MeV as a function of
the center-of-mass scattering angle.
The solid (dashed) line is the result with the nonlocal (localized) potential.
The results for $^{208}$Pb target are scaled by $10^{-3}$ for visibility.
}
\label{fig10}
\end{center}
\end{figure}

\begin{figure}[tbp]
\begin{center}
\includegraphics[width=0.45\textwidth,clip]{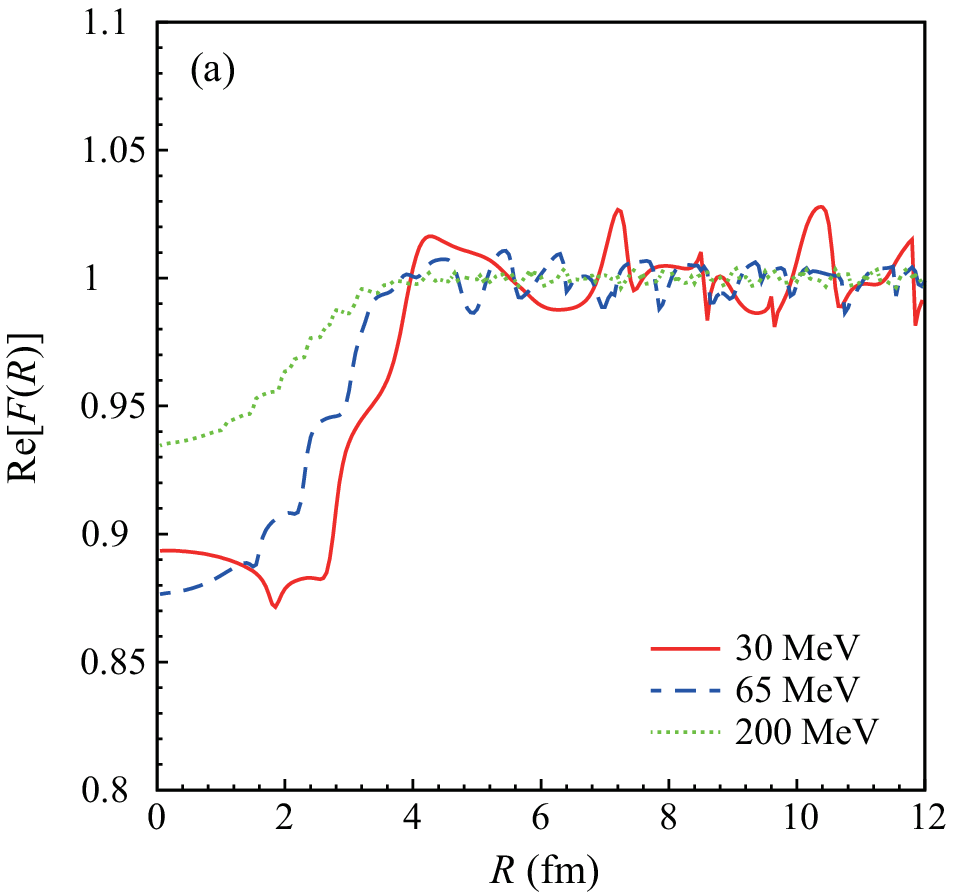}
\includegraphics[width=0.45\textwidth,clip]{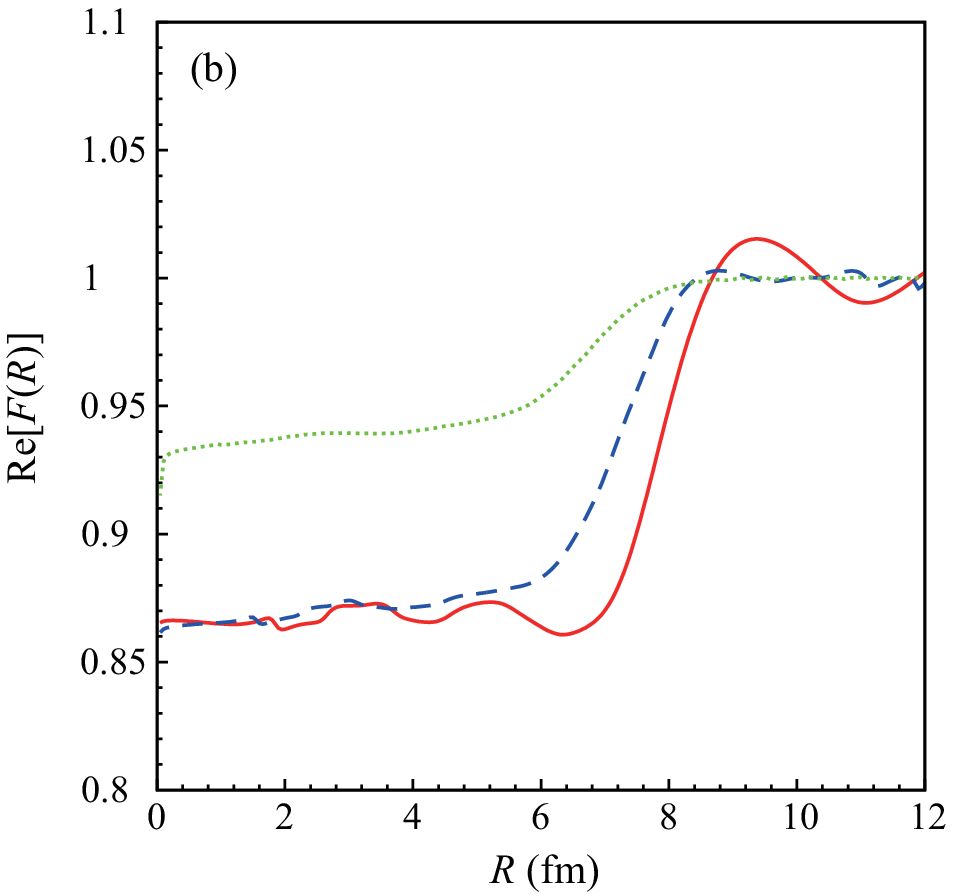}
\caption{(Color online)
Real part of the averaged Perey factor $F(R)$ for (a) $p$-$^{12}$C and (b) $p$-$^{208}$Pb scattering.
The solid, dashed, and dotted lines show the results at 30, 65, and 200 MeV, respectively.
}
\label{fig11}
\end{center}
\end{figure}

In general, microscopic optical potentials are nonlocal due to the antisymmetrization between the incoming and internal nucleons.
It was shown in Ref.~\cite{min10} the nonlocal effects can be well treated by a localization prescription proposed
in Refs.~\cite{BR77,BR77a,BR78}, that is, the Brieva-Rook (BR) localization, for proton elastic scattering on a $^{90}$Zr target.
In this study we examine the validity of the BR localization for the calculation of the elastic cross sections for $p$-$^{12}$C and $p$-$^{208}$Pb at 30~MeV (Fig.~\ref{fig10}),
$\sigma_{\rm R}$ for these at 30, 65, and 100~MeV (Table~\ref{tab1}),
and the scattering wave functions (Fig.~\ref{fig11}).
In the calculation, only the central part of the potential is included and we take $\vr_g=\vr$ as the density-dependence of the $g$-matrix interaction.
We use the target ground-state wave function obtained by the spherical Hartree-Fock calculation with the Gogny D1S interaction.

In Figure~\ref{fig10}, we show the cross sections for $p$-$^{12}$C and $p$-$^{208}$Pb at 30 MeV.
The solid (dashed) line represents the result with the nonlocal (localized) potential.
The BR localization is found to work rather well even at low energy.
Comparison on $\sigma_{\rm R}$ is shown in Table~\ref{tab1}, which justifies
the BR localization for also $\sigma_{\rm R}$ with error about a few percent.
\begin{table}[tbh]
\caption{
Proton total reaction cross sections $\sigma_{\rm R}^{}$ calculated with nonlocal and localized potentials.
}
\begin{center}
\begin{tabular}{c|c|c} \hline\hline
                         & \multicolumn{2}{c}{$\sigma_{\rm R}^{}$ (mb)} \\
                         & nonlocal & local    \\ \hline
 $p$-$^{12}$C@30MeV      & 441.0    & 452.2    \\
 $p$-$^{12}$C@65MeV      & 313.5    & 318.6    \\
 $p$-$^{12}$C@100MeV     & 241.9    & 244.6    \\ \hline
 $p$-$^{208}$Pb@30MeV    & 1697.9   & 1758.2   \\
 $p$-$^{208}$Pb@65MeV    & 1925.3   & 1945.2   \\
 $p$-$^{208}$Pb@100MeV   & 1758.2   & 1766.7   \\ \hline\hline
\end{tabular}
\label{tab1}
\end{center}
\end{table}

It is known that because of the nonlocality of the distorting potential
the amplitude of the scattering wave $\chi^{\rm (nonlocal)}$ obtained
with a nonlocal potential is smaller than that of $\chi^{\rm (local)}$,
a solution with a localized one in the nuclear interior region.
This is called the Perey effect and phenomenologically taken into account
by multiplying $\chi^{\rm (local)}$ by the so-called Perey factor.
In this study we take the ratio of $\chi_L^{\rm (nonlocal)}$ to
$\chi_L^{\rm (local)}$, and see the radial dependence of the Perey effect
in a more direct manner; $\chi_L^{\rm (nonlocal)}$ ($\chi_L^{\rm (local)}$)
is a partial wave of $\chi^{\rm (nonlocal)}$ ($\chi^{\rm (local)}$) with the
angular momentum $L$ between proton and the target nucleus.
To smear the $L$ dependence, we define an averaged Perey factor by
\bea
F(R)=\frac{1}{L_{\rm max}+1}\sum_{L=0}^{L_{\rm max}}
\frac{\chi_L^{\rm (nonlocal)}(R)}{\chi_L^{\rm (local)}(R)},
\eea
where $L_{\rm max}$ is the number for the partial waves.

Figure~\ref{fig11} shows the real part of $F(R)$ for $p$-$^{12}$C and $p$-$^{208}$Pb elastic scattering;
the imaginary part is almost zero since $\chi_L^{\rm (nonlocal)}(R)$ and $\chi_L^{\rm (local)}(R)$ are close to each other for all $L$.
One sees that $F(0)$, which can be regarded as a measure of the nonlocality,
does not depend strongly on the target nuclei. On the other hand, $F(0)$
shows clear dependence on the incident energy. This can be understood
by the fact that the exchange term of the interaction, the source of nonlocality in
the folding model calculation, has less contribution at higher energies.
The range of $F(R)$ depends on the target, reflecting the range of the exchange term.
The Perey effect shown in Fig.~\ref{fig11} must be included in reaction
calculations in which a scattering wave calculated with a localized potential is
involved, for example, transfer and knockout reactions. It should be noted
that $F(R)$ turns out to depend on NN effective interactions. More detailed
studies on $F(R)$ will be necessary.
Investigation on the nonlocality of the spin-orbit term will also be
important~\cite{kar17}.

\subsection{Deuteron scattering}
\label{sec3e}

\begin{figure}[tbp]
\begin{center}
\includegraphics[width=0.45\textwidth,clip]{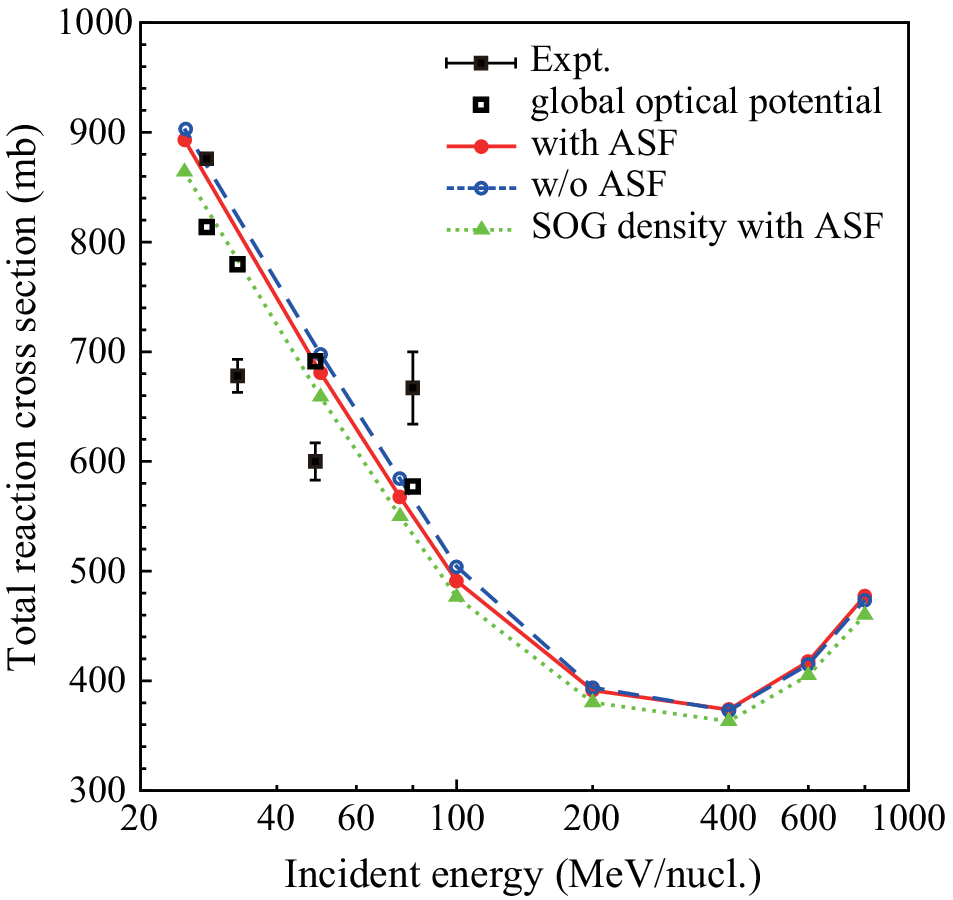}
\caption{(Color online)
Same as Fig.~\ref{fig5} but for $d$-$^{12}$C.
The experimental data are taken from Refs.~\cite{mat80,auc96,mil54}.
The results with the global optical potential for deuteron elastic scattering~\cite{an06} are shown by the open squares.
}
\label{fig12}
\end{center}
\end{figure}

We here investigate the effect of the ASF on deuteron scattering.
The model space of CDCC adopted in solving Eq.~(\ref{Schr_deu}) is the same as in Ref.~\cite{min17b}.
The effect on the deuteron elastic cross section is found to be very
similar to that on nucleon scattering discussed in Sec.~\ref{sec3a}.
Figure~\ref{fig12} shows the effect of the ASF on $\sigma_{\rm R}$ for $d$-$^{12}$C scattering.
The meaning of the symbols is the same as in Fig.~\ref{fig5}.
The experimental data are taken from Refs.~\cite{mat80,auc96,mil54}.
The results with the global optical potential for deuteron elastic scattering~\cite{an06},
which is applicable up to 91~MeV/nucleon, are shown by the open squares.
The effects of the ASF is rather small, because the factor is more close to unity than that for $p$-$^{12}$C.
There is a few percent difference between the SLy4-HFB and SOG densities.
However, both results are almost consistent with the measured data and the results with the global optical potential.

\section{Summary}
\label{sec4}
We have systematically examined the applicability of the microscopic
folding model based on the Melbourne $g$-matrix interaction and
the SLy4-HFB nuclear density.
Effect of the ASF appearing in multiple scattering theory, theoretical
uncertainty coming from the LDA, and the validity of the BR localization
are investigated in particular.
Deuteron-nucleus scattering is also considered to evaluate the
effect of the ASF on it; CDCC with nucleon-nucleus microscopic
potentials obtained by the folding model is adopted.

We found that the effect of the ASF is about 10\% on $\sigma_{\rm R}$
for $p$-$^{12}$C below 200 MeV resulting in better agreement with experimental
data, whereas it is negligibly small on $\sigma_{\rm R}$ in other cases
and on the elastic cross sections
in all the cases considered. The ASF should therefore be included in evaluating
the microscopic potential as well as the transition amplitude,
if the mass number of the target nucleus is small and the scattering energy
is low. At low energies, the LDA becomes less reliable and may cast
doubt on a $g$-matrix folding model and resulting observables.
The theoretical uncertainty regarding
the LDA was found to be about 10\% at most. One should keep this possible
uncertainty in mind when a $g$-matrix folding model is applied to
scattering at low energies. The BR localization turned out to work
quite well even below 65~MeV. The Perey effect on the scattering
wave was shown by directly comparing the scattering waves obtained
with nonlocal and localized microscopic potentials. Further investigation
on the Perey effect, its dependence on the effective NN interactions in
particular, will be necessary. Another important finding is that
the SLy4-HFB density can be an alternative to a phenomenological
density except for light nuclei, for example, $^{12}$C. This will
allow one to apply the framework to scattering of unstable nuclei
for which a phenomenological density is not available.

Thus, the microscopic folding model employing
the Melbourne $g$-matrix interaction and the SLy4-HFB density,
with the LDA and the BR localization implemented, was found to work
satisfactorily well
for describing nucleon-nucleus scattering at energies higher than
about 25~MeV. Possible deviation from experimental data will roughly be
10\%, depending on the reaction systems and observables.
A description of scattering at even lower energy will require
a new approach, having a wider model space and contains less approximations,
beyond the $g$-matrix folding model.

\section*{Acknowledgements}
The authors thank S. Karataglidis and K. Amos for valuable discussion.
The authors thank Y. S. Neoh for checking the numerical results partly.
This work is supported in part by
Grant-in-Aid for Scientific Research
(Nos. 16K05352 and 16K17698)
from Japan Society for the Promotion of Science (JSPS)
and by ImPACT Program of Council for Science,
Technology and Innovation (Cabinet Office, Government of Japan).
The numerical calculations in this work were performed at RCNP.


\end{document}